\documentclass[prb,aps,twocolumn,showpacs,preprintnumbers,amsmath,amssymb,superscriptaddress,unsortedaddress,floatfix]{revtex4}

\usepackage[dvips]{graphicx}
\usepackage{dcolumn}
\usepackage{bm}

\newcommand{\TN}{\ensuremath{T_{\mathrm{N}}}}

\renewcommand{\vec}[1]{\ensuremath{\bm{#1}}}

\begin{document}

\title{Resonant x-ray scattering study of NpRhGa$_5$ and NpCoGa$_5$}

\author{B. Detlefs}
\email{blanka.detlefs@esrf.fr}
\affiliation{%
European Commission, JRC, Institute for Transuranium Elements, 
Postfach 2340, Karlsruhe, D-76125, Germany}%
\affiliation{European Synchrotron Radiation Facility, BP 220, F-38043 
Grenoble, France}

\author{S. B. Wilkins}%
\altaffiliation[Present address: ]{Brookhaven National Laboratory, Condensed
Matter Physics and Materials Science Dept., Bldg \#510B
Upton, NY, 11973-5000, USA}
\affiliation{%
European Commission, JRC, Institute for Transuranium Elements, 
Postfach 2340, Karlsruhe, D-76125, Germany}%
\affiliation{European Synchrotron Radiation Facility, BP 220, F-38043 
Grenoble, France}

\author{R. Caciuffo}
\affiliation{%
European Commission, JRC, Institute for Transuranium Elements, 
Postfach 2340, Karlsruhe, D-76125, Germany}%

\author{J.A. Paix\~ao}
\affiliation{Departamento de Fisica, Universidade de Coimbra, P-3004 
516 Coimbra, Portugal}

\author{K.~Kaneko}
\affiliation{Advanced Science Research Center, Japan Atomic Energy 
Agency, Tokai, Ibaraki 319-1195, Japan}

\author{F. Honda}
\altaffiliation[Present address: ]{Dept. of Physics, Osaka University, Machikaneyama 1-1, Toyonaka,
560-0043, Japan}
\affiliation{Advanced Science Research Center, Japan Atomic Energy 
Agency, Tokai, Ibaraki 319-1195, Japan}

\author{N. Metoki}
\affiliation{Advanced Science Research Center, Japan Atomic Energy 
Agency, Tokai, Ibaraki 319-1195, Japan}

\author{N. Bernhoeft}
\affiliation{CEA Grenoble, DRFMC/SPSMS, F-38054, Grenoble, France}

\author{J. Rebizant}
\affiliation{%
European Commission, JRC, Institute for Transuranium Elements, 
Postfach 2340, Karlsruhe, D-76125, Germany}%

\author{G. H. Lander}
\affiliation{%
European Commission, JRC, Institute for Transuranium Elements, 
Postfach 2340, Karlsruhe, D-76125, Germany}%

\date{\today}

\begin{abstract}

We report a resonant x-ray scattering (RXS) study of antiferromagnetic 
neptunium compounds NpCoGa$_{5}$ and NpRhGa$_{5}$ at the Np $M_{4}$ and 
Ga $K$-edges. Large resonant signals of magnetic dipole character 
are observed below the N\'{e}el temperatures at both edges. The 
signals at the Np edges confirm the behaviour determined previously 
from neutron diffraction, i.e. the moments along $[001]$ in NpCoGa$_{5}$ 
and in NpRhGa$_{5}$ a reorientation of the moments from the $c$-axis 
direction to the $ab$ plane. In the latter material, on application 
of magnetic field of 9~Tesla along the $[010]$ direction we observe a change 
in the population of different $[110]$-type domains. We observe 
also a magnetic dipole signal at the Ga $K$-edge, similarly to the 
reported UGa$_{3}$ case, that can be interpreted within a semi-localized 
model as an orbital polarization of the Ga $4p$ states induced via strong hybridization 
with the Np $5f$ valence band. Quantitative analysis of the signal 
shows that the Ga dipole on the two different Ga sites follows 
closely the Np magnetic moment reorientation in NpRhGa$_{5}$. The 
ratios of the signals on the two inequivalent Ga sites are not 
the same for the different compounds.

\end{abstract}

\maketitle

\section{\label{sec:Introduction}Introduction}

The recent discovery of unconventional superconductivity in PuCoGa$_{5}$ 
and PuRhGa$_{5}$~\cite{Sarrao2002, Wastin2003} with relatively high superconducting 
temperatures $T_{c}\sim 18.5$~K and 8 K, respectively, 
has resulted in considerable interest in the electronic properties 
of materials with $5f$ electrons. Both experimental~\cite{Sarrao2002, 
Wastin2003, Curro2005} and theoretical~\cite{Opahle2003, Opahle2004, Shick2005, 
Pourovskii2006} studies indicate that magnetic interactions might 
be important for understanding the pairing mechanism in these 
compounds. Therefore, investigation of the related U and Np-115 
compounds is of interest with respect to the general properties 
of the actinide, $An$, and transition metal, $T$, isostructural $AnT$Ga$_{5}$. U$T$Ga$_{5}$ systems exhibit a variety of properties, 
ranging from Pauli paramagnetism in UCoGa$_{5}$~\cite{Sechovsky1992, Moreno2005, Troc2004}, 
to antiferromagnetism in UNiGa$_{5}$~\cite{Kaneko2003}. On the other hand, 
Np$T$Ga$_{5}$ compounds all show strong magnetic ordering~\cite{Colineau2004, 
Metoki2005,  Jonen2006a, Jonen2006, Honda2006, Honda2006a, Metoki2006}.

\begin{figure}[htb]
\includegraphics*[width=0.8\columnwidth]{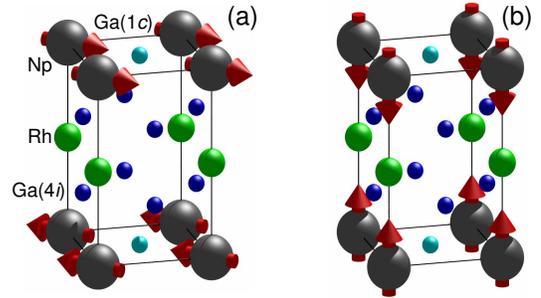}
\caption {\label{fig1} (Color online) Crystal and magnetic structure of NpRhGa$_{5}$ at $T < T^*$
(panel $a$) and at $T^* <T<\TN$ (panel $b$)~\cite{ Jonen2006a}). The magnetic structure of NpCoGa$_{5}$ is identical 
to the one of NpRhGa$_{5}$ at $T^* <T<\TN$ (panel $b$)~\cite{Metoki2005}.}
\end{figure}

NpCoGa$_{5}$ and NpRhGa$_{5}$ crystallize in the tetragonal HoCoGa$_{5}$-type 
structure which belongs to the space group $P4/mmm$ (no. 123), similarly to other actinide-based ``115'' compounds, 
including PuCoGa$_{5}$, see Fig.~\ref{fig1}. In this structure, actinide 
atoms occupy the $1a$ positions, transition metal ions  
are in the $1b$ positions (half-way between the actinide atoms 
along the $c$-direction) and there are two crystallographic 
positions for Ga: one atom in the center of the basal planes 
($1c$, addressed as Ga(1) in this paper), and 4 atoms in 
the $4i$ position, in the rectangular faces of the unit cell 
(Ga(2)) with a position (0 1/2 $z$). The Ga(1) site has 4 nearest 
Np neighbors, whereas the Ga(2) site has only two. For NpCoGa$_{5}$, 
with $z = 0.3103$~\cite{Colineau2004}, the nearest Ga neighbours of 
Np are at 2.964 {\AA} (Np -- Ga(2)) and 2.997 {\AA} (Np -- Ga(1)). 
In NpRhGa$_{5}$, with $z = 0.2987$~\cite{Colineau2005}, the Np -- Ga interatomic 
distances are very close for Np -- Ga(1) and Np -- Ga(2), namely 
2.964 and 2.963 {\AA}, respectively. 

For a diffraction experiment the crystal symmetry has an important 
feature: taking into account the magnetic propagation vector $\vec{q} 
= [0\ 0\ 1/2]$, a general magnetic diffraction peak of the type 
$(H, K, L\pm1/2)$ will always have a contribution 
from Ga(1) position, but Ga(2) will contribute only to reflections 
with $(H+K) = $ even, allowing the possibility to observe 
Ga(1) signal separately at $(H+K) = $ odd.

Resonant x-ray scattering (RXS) is a technique complementary 
to neutron diffraction for determining the magnetic structure 
of materials. In the case of actinides, the use of RXS presents 
an advantage over neutrons as significantly smaller samples can 
be studied, minimizing the safety restrictions imposed by the 
radiotoxicity of actinide isotopes. 

In the RXS process, a photon is virtually absorbed by a core level electron, which is promoted into an empty valence band state. The virtual excited state (core hole + photoelectron) then decays under emission of the scattered photon. The sensitivity of RXS to magnetism and orbital ordering arises from the periodic variation of the density of states near the Fermi level that are probed by the virtual photoelectron ~\cite{Blume1985, Gibbs1988}.
The intensity 
of the superlattice reflections increases dramatically as the 
photon energy is tuned to the atomic absorption edge of the resonant ion. Information 
can be obtained by measuring the polarization-dependent scattering 
intensity at forbidden Bragg positions as a function of photon 
energy and azimuthal rotation angle (the angle describing the 
rotation of the crystal about the scattering vector). Direct 
information on the magnetic polarization of the Np $5f$ states 
is obtained at the Np $M_{4}$ and $M_{5}$-edges, as electric 
dipole transitions involve the promotion of a $3d$ core electron 
to an empty $5f$ state. Dipole transitions at the Ga 
$K$-edge, on the other hand, involve the promotion of a $1s$ core 
electron of Ga to an empty $4p$ state.

Resonant scattering intensities with magnetic dipole symmetry 
at the $K$-edge of nominally nonmagnetic ions, such as Ga, 
have been reported previously for uranium compounds where the U sublattice 
is magnetically ordered~\cite{Mannix2001, Kuzushita2006, Normile2007}. The 
effect is related to the large spatial extent of $4p$ states 
within a broad energy band around the Fermi level making them sensitive 
to the electronic structure at neighboring sites. Although the 
$4p$ states of Ga are not spin-polarized, orbital polarization 
can be induced by hybridization with the U $5f$ states~\cite{Veenendaal2003, 
Usuda2004}. To our knowledge, no previous reports have been made 
of a similar phenomenon in a Np compound. Since, in general, 
increased localization is anticipated when the number of $5f$ 
electrons increases, this is not an obvious result.

\section{\label{sec:Methods}Experimental methods}

Single crystals of NpRhGa$_{5}$ and NpCoGa$_{5}$ were grown by a Ga 
self-flux method at the Institute for Transuranium Elements, 
in Karlsruhe. Samples with $(100)$ and $(001)$ surfaces 
were cut for the experiments and encapsulated in a copper container 
with a Be window for x-ray optical access.

Experiments were carried out on the magnetic scattering beamline 
ID20 at the European Synchrotron Radiation Facility, Grenoble, 
France. By tuning the undulators to their first or third harmonic, 
both Np $M_{4}$-edge ($E = 3.846$~keV) and Ga $K$-edge ($E 
= 10.364$~keV) were reached. A liquid nitrogen cooled Si(111) double 
monochromator and two vertically focusing Si mirrors provided 
a beam of 0.4 x 0.4 mm$^{2}$ at the sample position~\cite{Paolasini2007}.

\begin{figure}[htb]
\includegraphics*[width=0.8\columnwidth]{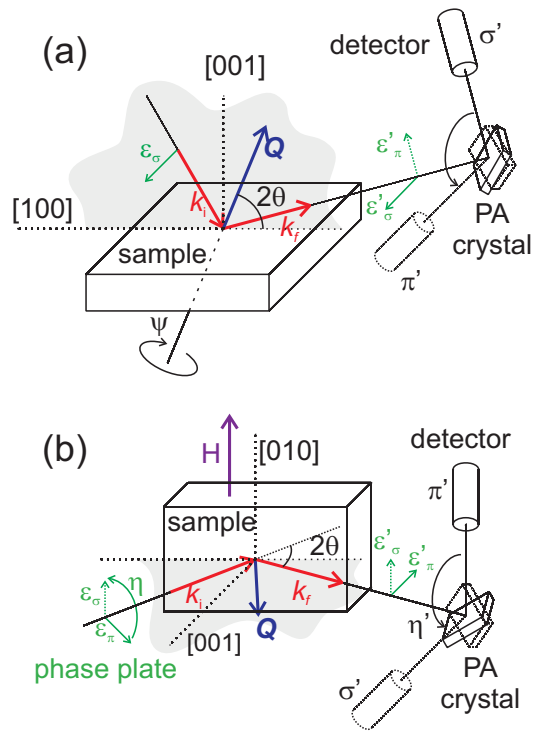}
\caption {\label{fig2} (Color online) Coordinate system and polarization vectors relative to the incident 
and scattered beams.}
\end{figure}

Two distinct scattering geometries and different sample environments 
were used: for measuring the azimuthal dependence 
of the superlattice peaks intensity, the sample capsule was mounted 
in a $^{4}$He closed-cycle cryostat with base temperature 
of about 12 K. A vertical scattering geometry on a 5-circle diffractometer 
with incident $\sigma$ x-ray polarization was used (see Fig.~\ref{fig2}a). In an azimuthal scan the sample rotates about the scattering 
vector, which implies a change of the angle between the incident 
electric field and the crystal axes. The resulting intensity 
oscillations give information on the symmetry and the orientation 
of the scattering tensor. In the vertical geometry, the instrumental 
setup does not support a heavy sample environment. The high magnetic 
field study on NpRhGa$_{5}$ was therefore performed in the second 
experimental hutch of ID20, using a horizontal geometry six-circle 
diffractometer equipped with a 10~T vertical split-pair superconducting 
magnet. In this geometry (Fig.~\ref{fig2}b), the incident light polarization 
is $\pi$, i.e. with the electric field vector lying in the scattering 
plane. Due to the restrictions imposed by the 10~T cryomagnet, 
an azimuthal scan is not possible. A polarization scan, giving similar information as the azimuthal one, can 
be performed using a phase-plate technique: instead of rotating 
the sample, the incident light polarization $\eta$ is turned from $\pi$ 
to $\sigma$: a diamond phase plate of thickness $100 \mu$m 
with a $[110]$ surface and the $(111)$ Bragg reflection can be operated 
in the half-wave plate mode so that the incident linear polarization 
can be rotated into an arbitrary plane by rotating the diamond 
crystal about the incident beam direction~\cite{Bouchenoire2006, Mazzoli2007}.

In all the scattering geometries, polarization analysis of the 
scattered beam was performed using the $(111)$ reflection from 
a Au crystal for data at the Np $M_{4}$-edge and pyrolitic graphite $(008)$ at the Ga $K$-edge. 
These crystals, acting as polarization filters, were chosen because of their $d$-spacings ($d_{(111)}^{\mathrm{Au}} = 2.3454$~\AA,
$d_{(008)}^{\mathrm{PG}} = 0.83860$~\AA) that 
give  Bragg diffraction close to the angle of $45^\circ$ 
at the corresponding energies.

When the phase plate technique is used in the polarization scan, full polarization analysis of the 
scattered beam is necessary in order to determine the Stokes 
parameters $P'_{1}$ and $P'_{2}$ by fitting the measured intensity 
to the function
\begin{equation}
I'=\frac{I'_{0} }{2} \left[ S+P'_{1} \cos \eta '+P'_{2} \sin \eta '\right] 
\label{eq:stokes_function}
\end{equation}
where the angle ${\eta}'$ describes the rotation of the polarization 
analyzer around the scattered beam and $S$ is the leakage, 
or component from the other polarized state. This arises due 
to the difference between the actual Bragg angle on the polarization 
analyzer and the ideal value of~$45^\circ$. 

From another point of view, $P'_{1}$ 
and $P'_{2}$ can be defined as
\begin{equation}
P'_{1} =\frac{|E'_\pi|^2-|E'_\sigma|^2}{|E'_\pi|^2+|E'_\sigma|^2}
\label{eq:P1}
\end{equation}
and 
\begin{equation}
P'_{2} =\frac{|E'_\pi+E'_\sigma|^2-|E'_\pi - E'_\sigma|^2}{2(|E'_\pi|^2+|E'_\sigma|^2)}
\label{eq:P2}
\end{equation}
where $E'_\sigma$ and $E'_\pi$ are the two components of the electric field vector of
the scattered photons~\cite{Born1999}. $P'_{2}$ is a measure 
of phase relation between $\sigma'$  and $\pi'$ signals.

However, the polarization analysis is not complete as only the $P'_{1}$ 
and $P'_{2}$ components can be measured in this way, but a distinction 
between a depolarization and an occurrence of a circularly polarized 
component $P'_{3}$ can be made on the basis of experimental data 
and simulations, because the degree of linear polarization $P'_\mathrm{lin}
= \sqrt{(P'_{1})^{2}+(P'_{2})^{2}}$ is closely related to the 
total beam polarization $P' =\sqrt{(P'_{1})^{2}+(P'_{2})^{2}+(P'_{3})^{2}}$. 
A real x-ray beam will have $P < 1$ (due to the finite 
efficiency of the phase plate) but it stays constant during a 
phase plate scan.

\section{\label{sec:Resonance}Resonance effects}

For better understanding of the magnetism in these compounds, searches for RXS signals were
performed at the Np $M_{4,5}$
 and Ga $K$-edges in  both NpCoGa$_{5}$ and 
NpRhGa$_{5}$. Resonant signals were found at positions in the reciprocal 
space corresponding to the magnetic propagation vector $\vec{q} 
= [0\ 0\ 1/2]$, i.e. ordering wavevectors previously found by neutron diffraction~\cite{Metoki2005,  Jonen2006a}. In both compounds, all resonant signals were  $\sigma\pi'$ polarized, indicating that they arise from magnetic dipole order.

In NpCoGa$_{5}$, resonances 
were found both at Np $M_{4}$ and $M_{5}$-edges, with the former 
much stronger and with a maximum at $3.846$~keV. Further data on 
the Np sublattice were recorded at this energy.
Superlattice diffraction peaks corresponding to the propagation $\vec{q} 
= [0\ 0\ 1/2]$ were found also at Co $K$-edge ($7.709$~keV) but 
the energy dependence of the signal suggests that it is non-resonant 
in character.

\begin{figure}[htb]
\includegraphics*[width=0.8\columnwidth]{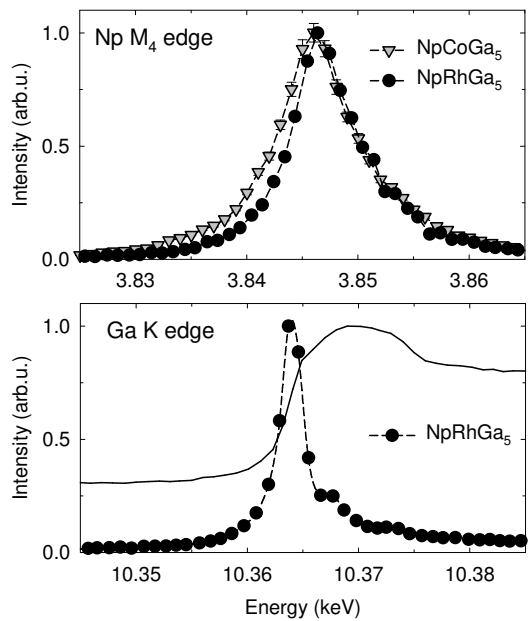}
\caption {\label{fig3}
Energy dependence of the resonant x-ray scattering signal in 
NpCoGa$_{5}$ and NpRhGa$_{5}$. Upper frame: Np $M_{4}$ data collected 
at $\vec{Q} = (0\ 0\ 2.5)$ (for NpCoGa$_{5}$) and $\vec{Q} = (0\ 0\ 
3.5)$ (for NpRhGa$_{5}$). In the case of NpCoGa$_{5}$, the intensity 
corresponds to integration over $\theta$ scans, in the case 
of NpRhGa$_{5}$ only energy scans with fixed $\vec{Q}$ were performed. 
Lower frame: Ga $K$-edge data recorded at $\vec{Q} = (0\ 0\ 6.5)$ 
for NpRhGa$_{5}$. All data were collected at base temperature ($T 
= 11.7$~K). The solid line in the lower panel represents the fluorescence 
spectrum at the Ga $K$-edge in NpRhGa$_{5}$. }
\end{figure}

The Np $M_4$ resonant spectra from both NpCoGa$_{5}$ and NpRhGa$_{5}$ are shown in the top 
panel of Fig.~\ref{fig3}. The slightly different widths of the two 
curves are due to the different mode of data collection. 
The spectral shape of the curve for the Np $M$-edges appears 
much as found in previous studies, a single Lorentzian signal 
with a full-width at half maximum (FWHM) of $\sim 8$~eV~\cite{Longfield2002a} 
showing the resonances are, as expected, E1-F$^{[1]}$ in nature~\cite{Hill1996}.

The Ga $K$-edge resonance is also similar to those found for 
uranium compounds~\cite{Mannix2001, Kuzushita2006}, centered close to 
the the absorption edge, but with perhaps a small ``bump'' on the 
high-energy side. The energy width of the resonance in the Co compound was \textit{larger} 
than in the Rh one, by about 50\%. This is a similar situation 
to that reported in UNiGa$_{5}$ and UPdGa$_{5}$~\cite{Kuzushita2006}, 
where the $3d$ compound (UNiGa$_{5}$) has a larger width at the 
Ga $K$-edge than the compound with the $4d$ transition-metal 
ion (UPdGa$_{5}$). As the magnetic structures of the two U$T$Ga$_5$ compounds differ, the effects are more difficult to compare. However, taken together with the present data, they provide rare direct evidence for the influence of the transition metal on the valence bands and the magnetic exchange coupling.

\section{\label{sec:Experiments}Experiments using the resonances}

The large resonant enhancement at the Ga $K$-edge observed 
in U$T$Ga$_{5}$ compounds~\cite{Kuzushita2006} is present also in Np$T$Ga$_{5}$ 
system, giving us the possibility to study, indirectly, the hybridization 
between the Np and Ga sublattices. Furthermore, the crystal structure 
of these compounds allows us to separate signal originating from 
two crystallographically inequivalent Ga positions in the lattice 
by examining different reflections with $(H+K)$ even and odd, 
which was not done in the case of the U$T$Ga$_{5}$ systems.

\subsection{\label{subsec:NpCoGa5}NpCoGa$_5$}

\begin{figure}[htb]
\includegraphics*[width=0.8\columnwidth]{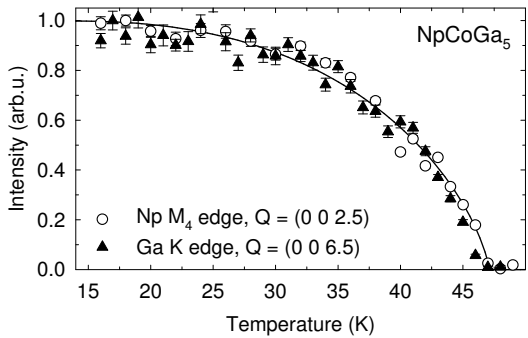}
\caption {\label{fig4}
Temperature dependence of the RXS signal in NpCoGa$_{5}$ at $\vec{Q} 
= (0\ 0\ 2.5)$ at Np $M_{4}$-edge (open symbols) and at $\vec{Q} 
= (0\ 0\ 6.5)$ at Ga $K$-edge (filled symbols). The solid line 
serves as a guide for the eye.}
\end{figure}

Fig.~\ref{fig4} shows a comparison of the temperature dependence of 
Np $M_{4}$ and Ga $K$-edge signals in NpCoGa$_{5}$ in the rotated 
polarization channel $\sigma \pi'$. In both cases a specular 
magnetic Bragg reflection $(0\ 0\ L+1/2)$ is shown. For the Ga 
$K$-edge this type of reflection includes contribution from the two 
crystallographically inequivalent Ga positions. The match in 
the temperature dependence of the two resonant signals indicates 
a common origin of the scattering process associated with the 
antiferromagnetic (AF) state. The N\'{e}el temperature $\TN$ 
of $46.9(2)$~K deduced from our measurements is in good agreement 
with the previous bulk measurements~\cite{Colineau2004} and neutron 
diffraction data~\cite{Metoki2005}. The Ga $K$-edge data were collected 
with the primary beam attenuated by $\sim 300 \mu$m aluminium 
foil corresponding to transmission of about 15\%. Such an attenuation 
was found necessary to avoid local beam heating of the sample 
at the Ga $K$-edge. 

\begin{figure}[htb]
\includegraphics*[width=0.8\columnwidth]{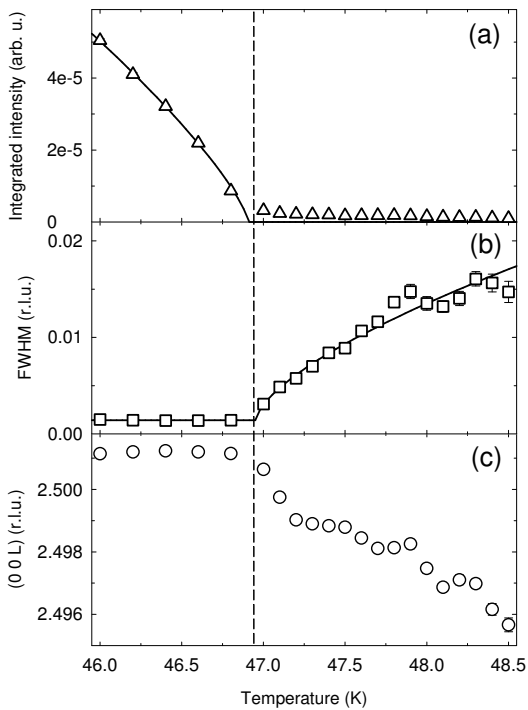}
\caption {\label{fig5}
Details of the temperature dependence of the RXS signal at the Np $M_4$-edge in NpCoGa$_{5}$:
(a) integrated intensity, (b) full-width at half-maximum, and (c) reciprocal space
position of the $\vec{Q} = (0\ 0\ 2.5)$ magnetic Bragg peak.
The solid lines are fit to the data giving the $\beta$ (upper panel)
and $\nu$ (middle panel) critical parameters. $\TN$ is marked as a dashed vertical line.}
\end{figure}

The temperature dependence of the Np $M_4$-edge signal in NpCoGa$_{5}$ 
was measured in more detail at $\vec{Q} = (0\ 0\ 2.5)$ (see Fig~\ref{fig5}).
Figure~\ref{fig5}(a) shows the temperature dependence of the integrated 
intensity. From a parametric fit of the scaling law for magnetization
$I \propto M^2 \propto (-t)^{2\beta}$, where $t = (T - \TN)/\TN$, we deduce a value of $\beta = 0.39(1)$, 
which is in good agreement with the value of $0.37(1)$ obtained from 
neutron scattering~\cite{Metoki2005}. Figure~\ref{fig5}(b) shows the variation 
of the FWHM, which gives the critical exponent $\nu = 0.66(8)$ 
relating the thermal variation of the correlation lengths in 
the paramagnetic state. Theoretical values for $\beta$ and $\nu$ are $0.37$ and $0.71$ 
for the 3D Heisenberg antiferromagnet, and $0.33$ and $0.63$ for 
a 3D Ising system~\cite{Collins1989}. Although our data on $\nu$ 
are not of sufficient accuracy to make a definitive statement, 
our experiments give little evidence for any special magnetic 
anisotropy (as would be the case for a 3D Ising system) in these 
materials.

The final frame, Fig.~\ref{fig5}(c) shows the change in the position of 
the magnetic diffraction peak. The small offset from 2.500 below $\TN$ arises from a systematic error in the determination of the lattice parameter. As verified by measuring the charge $(0\ 0\ 2)$ reflection, this systematic error stays constant with temperature, so that the variation of the position between $\TN$ and 48.5~K must be considered real even if it is only 2 parts in 
$10^{3}$. This shift of the peak in the paramagnetic state is not 
connected with the development of incommensurate fluctuations, 
otherwise there would be a matching peak from the $(0\ 0\ 3)$ charge 
peak seen just \textit{above} $(0\ 0\ 2.5)$. Instead, this is another 
observation of the so-called $\vec{q}$-shift~\cite{Bernhoeft2004a}. Given 
the extremely good resolution of synchrotron x-rays, especially 
at energies as low as the Np $M$-edges, this is relatively 
easy to observe with good crystals, but is not yet understood.

Turning to the Ga resonance, the focus is on the direction 
of the apparent dipoles at the Ga sites and their relative magnitudes. 
Since in NpCoGa$_{5}$ the Np moments are along the high symmetry 
direction of the structure, the $c$-axis, at all temperatures, 
it is expected that any effective Ga dipole will be along this direction,
as has been inferred by NMR/NQR 
experiments at the Ga sites~\cite{Kambe2007, Sakai2007}.

For the direction of the dipoles, the most informative measurements 
in RXS experiments are those tracing the azimuthal dependence 
of the intensity in both $\sigma \sigma'$ and $\sigma \pi'$ polarization channels~\cite{Hill1996, Paixao2002, 
Wilkins2006}.

\begin{figure}[htb]
\includegraphics*[width=0.8\columnwidth]{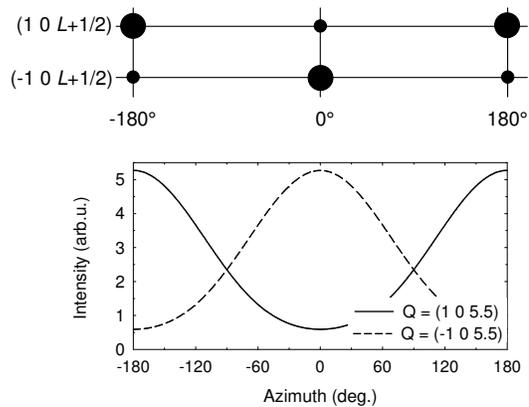}
\caption {\label{fig6}
Schematic representation of the intensity of azimuthal scans 
at reflections $\vec{Q} = (\pm1\ 0\ L+1/2)$ for NpCoGa$_{5}$ taken 
at the Ga $K$-edge. Only Ga(1) contributes at this scattering 
vector. Large symbols in the top panel correspond to high measured 
intensity; small circles correspond to low intensity. Lower panel: 
A simulated azimuthal scan for $\vec{Q} = (1\ 0\ 5.5)$ and $\vec{Q} 
= (-1\ 0\ 5.5)$ with effective Ga dipole moment parallel to $[001]$.}
\end{figure}

The Ga $K$-edge resonant signal was present only 
in the $\sigma \pi'$ scattering channel suggesting a magnetic 
dipole-like origin of the scattering, as found previously for 
signals at the Ga $K$-edge~\cite{Mannix2001, Kuzushita2006}. Collecting 
azimuthal scans in order to determine the direction of the dipoles 
on either of the Ga sites turned out to be experimentally 
difficult due to multiple scattering effects interfering with 
the signal. Such effects manifest themselves as spikes of high 
intensity in the data for certain values of azimuth. Despite 
these experimental artefacts a consistent picture may be given in which
a schematic representation of the data, 
together with a simulated azimuthal dependence is shown in Fig.~\ref{fig6}.
The observed intensities are represented in the top frame 
by large (small) points for high (low) intensity for the azimuthal 
angles shown. This pattern of intensities was found for a series 
of reflections with different $L$ values. The lower frame of Fig.~\ref{fig6}
shows the intensity of the $\vec{Q} = (\pm 1\ 0\ 5.5)$ reflections, calculated
for an effective dipole moment along $[001]$, as an example. 
In general, for a chosen $\vec{Q} = (H\ 0\ L)$ reflection 
and the actual scattering geometry (see Fig.~\ref{fig2}a) the azimuthal 
dependence for a signal originating from magnetic dipoles oriented 
along the $[0 0 1]$ direction can be analytically written as
\begin{equation}
I_{\sigma \pi '} \propto \left| f_{\sigma \pi '} \right| ^{2} \propto
\left| \frac{\sin \theta _{(H0L)} +\delta \cos \theta _{(H0L)} \sin \psi
}{\sqrt{1+\delta ^{2} } } \right| ^{2}
\label{eq:az_dep}
\end{equation}
where $\theta_{(H 0 L)}$ is the Bragg angle for 
the $\vec{Q} = (H 0 L)$, $\psi$ is the azimuthal 
value and $\delta$ is a parameter characterizing the off-specular nature 
of the chosen reflection, 
$
\delta =\frac{Ha^*}{Lc^*}.
$
This means that, when going to higher values of $L$ at fixed $H$, the difference 
between the $0^\circ$ and $\pm 180^\circ$ changes less dramatically 
on approaching the $[0 0 1]$ direction, parallel to the 
direction of the effective Ga magnetic dipoles.

For a fixed azimuthal angle, in this case $\psi = 0$, we 
have made a study of the intensities at the Ga $K$-edge for 
a series of reflections, both along the specular $(0\ 0\ L)$ 
and off-specular $(1\ 0\ L)$ directions. A comparison of these 
intensities allows a determination of the dipole signals associated 
with the Ga(1) and Ga(2) sites to be made. Recall that, because 
of the symmetry, Ga(1) contributes to \textit{all} reflections, and 
Ga(2) only to those with $H+K=$~even. 

\begin{table*}[hbt]
\begin{ruledtabular}
\begin{tabular}{c|cccccc|c|c}
& \multicolumn{6}{c|}{calculations} & exp. &  \\
\hline
$\vec{Q}$ & $L$ & $A$ & $f_\mathrm{Ga(1)}$ & $f_\mathrm{Ga(2)}$ & $t_{\sigma \pi'}$ & 
$I_{\sigma \pi'}$ & $I_\mathrm{exp}$ & $I_\mathrm{calc}/I_\mathrm{exp}$ \\
\hline
$(1\ 0\ 5.5)$ & 1.145 & 0.2243 & -2 & 0 & 0.4847 & 0.2419 & 2.4(3) & 0.10(1) \\
$(1\ 0\ 6.5)$ & 1.050 & 0.2786 & -2 & 0 & 0.5728 & 0.3839 & 3.2(4) & 0.12(2) \\
$(1\ 0\ 7.5)$ & 1.004 & 0.3279 & -2 & 0 & 0.6610 & 0.5752 & 5.0(5) & 0.11(1) \\
$(1\ 0\ 8.5)$ & 1.013 & 0.3748 & -2 & 0 & 0.7491 & 0.8525 & 6.0(5) & 0.14(1) \\
\hline
$(0\ 0\ 5.5)$ & 1.179  &0.2424  &2  &-2.152  &0.4847  &0.1933  &1.3(3)  &0.14(3)  \\
$(0\ 0\ 6.5)$  &1.065  &0.2864  &2  &7.955  &0.5728  &0.9753  &6.2(5)  &0.15(1)  \\
$(0\ 0\ 7.5)$  & 1.008  &0.3305  &2  &-3.732  &0.6610  &0.3161  &1.8(5)  & 0.17(5)  \\
$(0\ 0\ 8.5)$  &1.008  &0.3745  &2  &-5.194  & 0.7491  &0.3403  & 2.3(3)  & 0.14(2)  \\
\end{tabular}
\end{ruledtabular}

\caption {\label{tab1} Comparison of the calculated and observed intensities 
in NpCoGa$_{5}$. For details, see the text. The experimental uncertainties 
in parenthesis refer to the last significant digit. $I_{\sigma \pi'}$ is the calculated
intensity $I_\mathrm{calc}$.}
\end{table*}

Table~\ref{tab1} shows the comparison between the calculated and observed 
intensities of the resonant scattering signal in NpCoGa$_{5}$ at $T 
= 12$~K at the Ga $K$-edge. The calculated structure factors 
for both Ga positions are given $(f_\mathrm{Ga(1)}$ and $f_\mathrm{Ga(2)}$) 
as well as the resonant geometric factor $t_{\sigma \pi'}$~\cite{Hill1996}.
The calculated RXS intensity is corrected for the Lorentz 
factor $L = (\sin \theta)^{-1}$ and for an angular factor $A$ 
that accounts for the fraction of the incident beam intercepted 
by the sample
\begin{equation}
A=\frac{\sin (\theta +\alpha )\sin (\theta -\alpha )}{2\sin \theta \cos
\alpha }.
\end{equation}
Here, $\alpha$ is an ``asymmetry'' angle, 
i.e. the angle between the specular direction and the scattering 
vector~\cite{Detlefs1997}.

Given the often greater than 10\% uncertainty in the observed 
intensities, the agreement between the observed and calculated 
intensities is reasonable, although there does appear to be a 
different scale factor needed between the specular and off-specular 
reflections. This is almost certainly due to absorption effects 
not properly accounted for, especially as the surface of the 
crystal was not polished (i.e. flat), even though the crystal 
quality was good. The best agreement, as shown in the final column 
of the Table, for the $(1\ 0\ L+1/2)$ and $(0\ 0\ L+1/2)$ series 
of scattering vectors is for a ratio between Ga(1) and Ga(2) 
dipoles of:
\begin{equation}
m[\mathrm{Ga(2)}] = 0.14(2)\cdot m[\mathrm{Ga(1)}].
\end{equation}

\subsection{\label{subsec:NpRhGa5}NpRhGa$_5$}

The magnetic structure of NpRhGa$_{5}$ below $\TN = 37$~K is more 
complex than the one of NpCoGa$_{5}$ due to the magnetic phase transition 
occurring at $T^* = 32$~K where the Np magnetic moments re-orient 
from the $c$-axis direction to lie in the basal $ab$ plane 
(see Fig.~\ref{fig1}), with their moments along 
$[110]$ and $[1\bar{1}0]$ directions according to neutron scattering~\cite{ Jonen2006a}. 
This magnetic phase transition is accompanied by a discontinuous 
change in the magnetic moment amplitude suggesting a change of 
the $5f$ electronic states~\cite{Jonen2006a}.

In our RXS experiment we are able to test the conclusions of 
the neutron experiments in three different ways; first by examining 
the temperature dependencies of the intensities; second, by performing 
azimuthal scans; and third, by examining the polarization of 
the scattered photons to determine the domain population.

\begin{figure}[htb]
\includegraphics*[width=0.8\columnwidth]{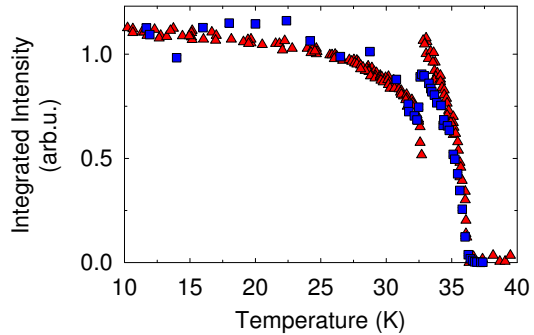}
\caption {\label{fig7} (Color online)
Comparison of the temperature dependence of the RXS with neutron-derived 
intensities [after Ref.~\onlinecite{Jonen2006a}] in NpRhGa$_{5}$. The RXS data (blue 
squares) from the $(0\ 0\ 3.5)$ reflection taken at the Np 
$M_4$-edge are shown unaltered; the neutron magnetic moment is squared 
to be comparable with the intensities and multiplied by the geometrical 
factors involved in the RXS experiment (red triangles). }
\end{figure}

Fig.~\ref{fig7} shows the temperature dependence of the RXS signal 
in NpRhGa$_{5}$ at the Np $M_4$-edge for $\vec{Q} = (0\ 0\ 3.5)$. 
The RXS signal is compared with the temperature dependence of 
the Np moment as deduced from neutron scattering by multiplying 
the moment (squared) by the relevant geometric scattering factors. Two effects 
are present in this simulation. The first is the growth of the 
moment as a function of temperature. This is taken from Fig. 
5 of Ref.~\onlinecite{Jonen2006a} and includes a discontinuity at $T^*$. The 
second is the re-orientation of the moments at $T^*$ that has 
a drastic effect on both the neutron and x-ray intensities. The 
good agreement between the neutron-derived and RXS intensity 
is the first indication that the envisaged model for the two 
magnetically ordered phases is correct.

The agreement shown in Fig.~\ref{fig7} is probably as good as can be expected 
between the two techniques, and shows that the re-orientation 
of the Np moments at $T^*$ away from the $[001]$ axis certainly 
occurs. Small differences between the neutron and RXS results 
around $T^*$ may arise from the lack of temperature stability in 
the RXS case, or, more interestingly, because the scale factor 
relating the moment to the RXS signal \textit{changes} at $T^*$. This 
might suggest a transition between two subtly different electronic 
states, as already suggested by Jonen \textit{et al.,}\cite{Jonen2006a} from their 
neutron experiments.

Although the moment direction in the high-temperature state,
$T^* < T < \TN$ is clearly established as $[001]$, the 
situation below $T^*$ is more complex as the moments can have 
any direction in the basal $ab$ plane. The neutron experiments 
are interpreted with the moments along $[110]$ and $[1\bar{1}0]$ directions, 
forming two domains. We therefore set out to confirm this using 
the new 10 T magnet facility available at ID20. Unfortunately, 
use of the heavy sample environment, such as the 10~T cryomagnet 
makes azimuthal scans impossible so that phase plate scans, described in Sec.~\ref{sec:Methods},
were necessary.

\begin{figure*}[bht]
\includegraphics*[width=1.7\columnwidth]{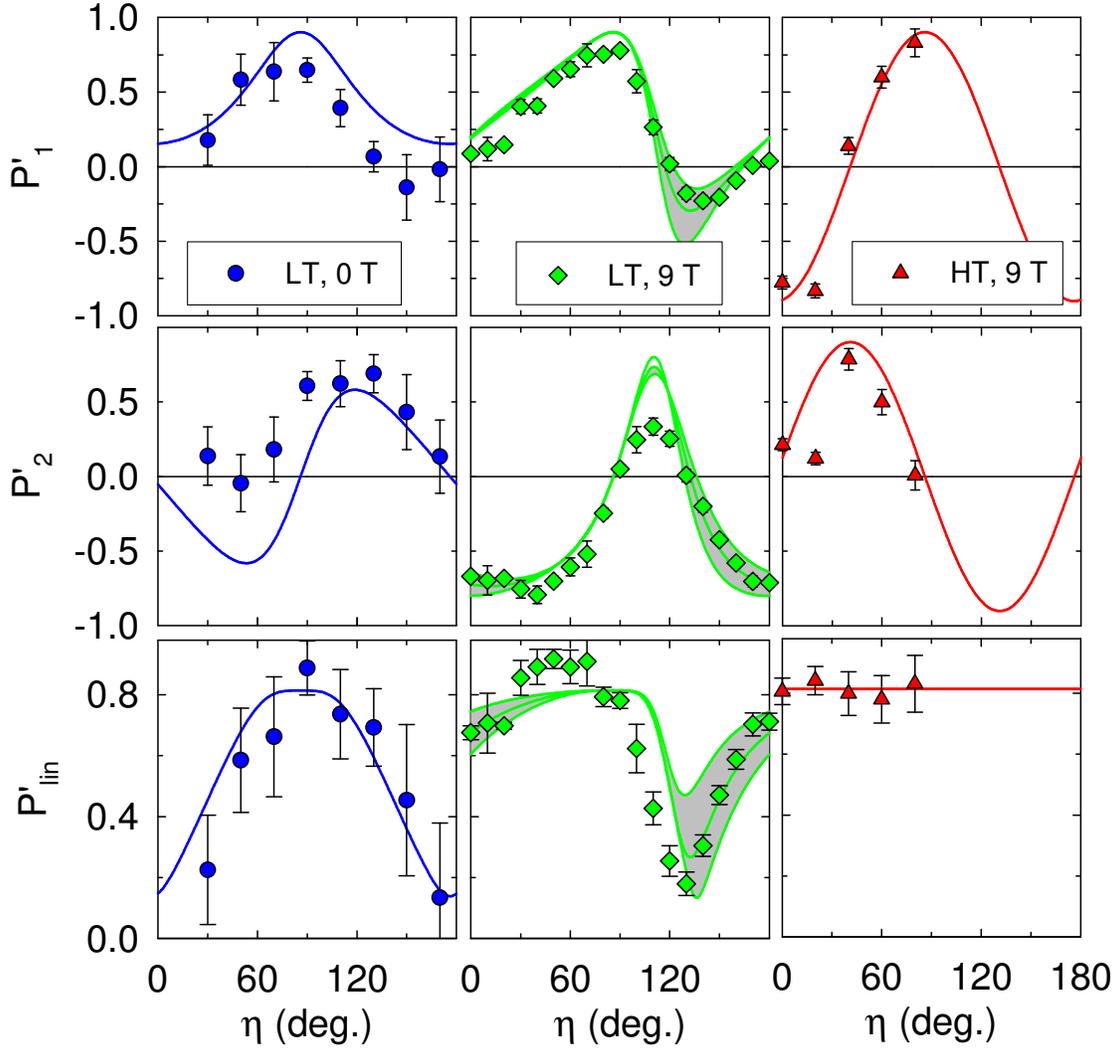}
\caption {\label{fig8} (Color online)
$P'_{1}$, $P'_{2}$ and $P'_\mathrm{lin}$ in the pseudo-azimuthal dependence 
(a phase-plate scan) of the Np $M_4$-edge signal in NpRhGa$_{5}$ 
at $\vec{Q} = (0\ 0\ 2.5)$ at $T = 20$~K (denoted as LT) and $33$~K
(HT) with magnetic field $B = 9$~T $\parallel b$-axis. In the scan, the incident light polarization
$\eta$ is being rotated from the $\sigma$ state ($\eta = 0^\circ$) through
the $\pi$ state ($\eta = 90^\circ$) back to the $\sigma$ state ($\eta =
180^\circ$). All the zero-field data are recorded in the zero-field cooled 
(ZFC) state. The solid lines correspond to the most probable model 
for the Np magnetic moments: equal population of the $[110]$ and 
$[1\bar{1}0]$ domains for ZFC data at $T = 20$~K, and approximately 
a 9:1 ratio for domain population at $B = 9$~T; the shaded area in the
central panels are delimited by simulation curves for domain populations
of 95:5 and 85:15, see text. The 
red curve corresponds to the $c$-axis moment model (HT 
phase). }
\end{figure*}

Fig.~\ref{fig8} shows $P'_{1}$, $P'_{2}$ and $P'_\mathrm{lin}$ values measured
at the Np $M_4$-edge in NpRhGa$_{5}$
as a function of polarization direction, $\eta$, of the incident light. The data were recorded 
at $\vec{Q} = (0\ 0\ 2.5)$ at $T = 20$~K (i.e. at $T < T^*$, 
denoted as LT) and 33 K (i.e. at $T^* < T < \TN$, denoted as HT) in both zero-field state and with 
magnetic field $B = 9$~T applied along the $b$-axis. Although 
the zero-field data were not measured with high statistics (the 
error bars correspond not only to the errors coming from the 
integration of the analyzer rocking curves but also from the 
fit to the Stokes function (Eq.~\ref{eq:stokes_function})) they agree with the model 
shown in blue solid line -- equal population of domains with Np 
magnetic moments along $[110]$ and $[1\bar{1}0]$ directions. Comparison 
of the zero-field data and the 9 T data in the $T < T^*$ 
(LT) phase (blue circles and green diamonds) shows the magnetic 
field effect: the domain population for the 9~T LT phase changes 
to 90\% $[110]$ domains vs. 10\% $[1\bar{1}0]$ domains, indicating that 
the applied magnetic field of 9~T is insufficient to induce 
a single domain, although it goes a long way to establish such 
a monodomain state. The choice of $[110]$ over $[1\bar{1}0]$ is a result 
of the field not being exactly along the $[010]$ direction, i.e. 
a small misorientation of the crystal.

The difference between the lower panels for the LT 
state shows that we are particularly sensitive to the direction \textit{and} 
the domain population with this technique. The difference in 
the values of $P'_\mathrm{lin}$ and 
the measured value $P = 0.81(2)$ is the result of \textit{depolarization} 
of the scattered photon beam, which can only occur if domains with different scattered beam
polarizations contribute to the total signal. For 
a single domain state (e.g. the HT state) there is no depolarization, 
and the linear scattered component is the same as the incident 
component, i.e. $P'_\mathrm{lin} = P$. It is thus interesting to observe the 
minima in $P'_\mathrm{lin}$. For 
an equal domain state, the minima occur at the symmetric positions $\eta
= 0^\circ, 180^\circ$, whereas as the domain population 
changes the minima in $P'_{\mathrm{lin}}$ change also. The shaded simulations 
are for models with domain populations of 95:5 and 85:15. The 
most probable domain population appears to be 90:10 at $B 
= 9$~T. The HT phase (represented by red triangles in Fig.~\ref{fig8}) 
data can be fit to the Np magnetic moments pointing along 
the $[001]$ direction even when magnetic field of 9 T is applied 
along the $b$-axis, which is consistent with the known 
phase-diagram in a magnetic field~\cite{Colineau2006}.

The sensitivity of the polarization of the scattered particle
to the moment direction and domain population is also exploited in spherical 
neutron polarimetry (SNP)~\cite{Brown2006}. In both cases the 
information may be obtained from a \textit{single} reflection, rather 
than the standard technique of comparing intensities of a number 
of different reflections~\cite{Hiess2001}. However, the x-ray technique 
may be used in a magnetic field, whereas the nature of the neutron 
interaction with a magnetic field restricts the SNP technique 
to zero applied field. Clearly, this is a considerable advantage 
when the domain population, as well as possibly the moment directions, 
can be changed by the application of a magnetic field.

\begin{figure}[htb]
\includegraphics*[width=0.8\columnwidth]{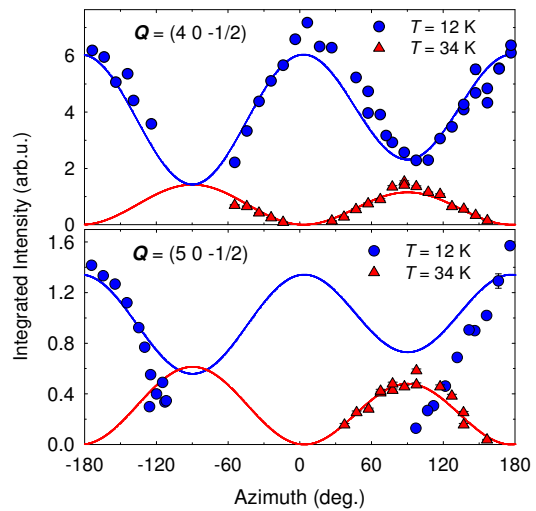}
\caption {\label{fig9} (Color online)
Azimuthal dependence of the Ga $K$-edge resonant signal ($\sigma\pi'$ 
channel) at two scattering vectors $\vec{Q} = (4\ 0\ -1/2)$ and $\vec{Q} 
= (5\ 0\ -1/2)$ at $T = 12$~K and 34~K. Solid lines correspond 
to the model described in the text. }
\end{figure}

Turning now to the Ga $K$-edge spectra, we show in Fig.~\ref{fig9} 
the results for two reflections in zero applied field but at 
different temperatures, above (triangles) and below (circles) 
the re-orientation transition at $T^*$. Only Ga(1) contributes 
to the signal at $\vec{Q} = (5\ 0\ -1/2)$ but contributions from 
both Ga sites are present at $\vec{Q} = (4\ 0\ -1/2)$. In the HT 
state ($T = 34$~K) the Ga dipoles are assumed to point along 
the $[001]$ axis and the agreement for both reflections with the 
model is acceptable. 

The intensities at the Ga $K$-edge clearly scale approximately 
as the moment on the Np site, which changes considerably below $T^*$, 
mostly because of the thermal effects as $\TN \sim 36$~K,
but there is also a discontinuity in the moment itself~\cite{Jonen2006a}. 
In the LT state there are some discrepancies with the model, 
but they are attributed to difficulties in measuring the intensity 
as a function of azimuthal angle over such a large azimuthal 
range. The domain population has been assumed as 50:50, corresponding to the 
ZFC state discussed in connection with Fig.~\ref{fig8}.

\begin{figure}[htb]
\includegraphics*[width=0.8\columnwidth]{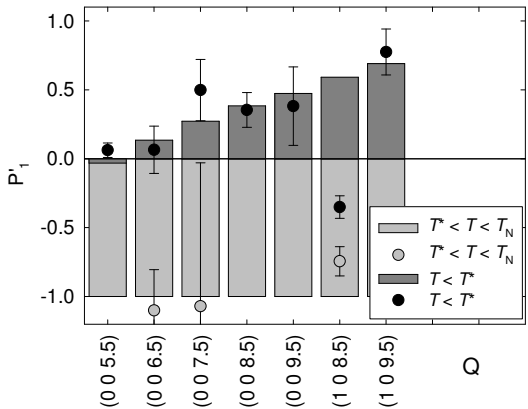}
\caption {\label{fig10} Stokes parameters analysis for NpRhGa$_{5}$ at the Ga 
$K$-edge. In the high-temperature phase $P'_1 = -1$ for \textit{all} 
reflections; the measured values are indicated with open points. 
In the low-temperature phase ($T< T^*$) the calculated values 
are given by the height of the bar graphs, with the measured 
values as closed circles. The measurements are for $B = 0$~T, so 
an equal domain population model is assumed. }
\end{figure}

The method of determining the Stokes parameters as used in Fig.~\ref{fig8}
for the Np $M_4$ resonance was used also for the Ga 
$K$-edge and can confirm the re-orientation of the Ga dipoles as 
suggested by the results and modelling in Fig.~\ref{fig9}. In the case 
of the Ga $K$-edge resonance there is a difficulty in 
making the integration of the analyzer crystal as strong inelastic 
scattering, associated with the $K\alpha_{1}$ and $K\alpha_{2}$ emission lines at 
about 1000 eV below resonance, can affect the measurement of 
the integrated intensities, and thus $P'_1$ and $P'_2$. 
 We show in Fig.~\ref{fig10} the results for $P'_1$ 
in the two states for a number of different reflections. 

In the HT state, $I_{\pi\pi'} = 0$ exactly 
for the specular reflections and it is almost zero for off-specular 
ones due to the azimuth value given by the direction of magnetic 
field with respect to the sample. From this and from the definition 
of $P'_{1}$ (Eq.~\ref{eq:P1}) it follows that $P'_1 = -1$ for all specular reflections and it 
is very close to $-1$ for off-specular ones. Although the $P'_1$ was not measured
at all reflections shown in Fig. 10, subsequent temperature dependence curves
on these reflections confirmed the absence of the $\pi \pi'$ channel signal.

At the moment reorientation temperature $T^*$ some of the intensity is transferred
into the $\pi \pi'$ channel so that in the LT state the sign of $P'_{1}$ changes and  
approaches $+1$, as the ratio $I_{\pi\pi'}/I_{\pi\sigma'}$
increases with increasing scattering angle $\theta_{(H0L)}$ and 
 decreasing off-specularity $\delta$. The agreement between 
the experimental and calculated $P'_{1}$ values indicate that the
direction of the moments on both 
Ga(1) and Ga(2) follow that of the  Np magnetic moment. Recall that both Ga sites 
contribute to the reflections with $(H +K) =$~even, but 
only the Ga(1) to reflections with $(H + K) =$~odd.

Similarly to the case of NpCoGa$_5$, intensities of several reflections must be compared in order to  estimate the ratio between amplitudes of the Ga(1) to Ga(2) dipoles. Table~\ref{tab2} shows this comparison in the LT phase, i.e. when model with dipoles pointing along the $[110]$ directions and an equal domain population is considered. The calculated structure factors for both Ga positions are given ($f_\mathrm{Ga(1)}$ and $f_\mathrm{Ga(2)}$) as well as the resonant geometric factor $t_{\sigma \pi'}$, calculated for the azimuthal angle $\psi = -86.06^\circ$ with respect to the $[010]$ azimuthal reference. The best agreement, as shown in the final column of the Table,  gives 
\begin{equation}
m[\mathrm{Ga(2)}] = 0.43(4)\cdot m[\mathrm{Ga(1)}].
\end{equation}

A similar analysis in the HT phase was not performed, because of difficulties due 
to the limited stability temperature range of this phase.

\begin{table*}[htb]
\begin{ruledtabular}
\begin{tabular}{c|ccccccc|c|c}
& \multicolumn{7}{c|}{calculations} & exp. &  \\
\hline
$\vec{Q}$ & $L$ & $A$ & $f_\mathrm{Ga(1)}$ & $f_\mathrm{Ga(2)}$ & $t_{\pi \sigma'}(110)$ & 
$t_{\pi \sigma'}(1\bar{1}0)$ & $I_{\pi \sigma'}$ & $I_\mathrm{exp}$ & $I_\mathrm{calc}/I_\mathrm{exp}$ \\
\hline
$(0\ 0\ 6.5)$  &1.065  &0.2864  &2  &7.467  &-0.618	&-0.538	&2.782	&54(1)	&19.4(4)  \\
$(1\ 0\ 7.5)$ & 1.004 & 0.3279 & -2 & 0 & -0.448	&-0.376	&0.225	&4.1(2)	&18(1) \\
$(1\ 0\ 9.5)$ & 1.113 & 0.4200 & -2 & 0 & -0.296 &-0.245	&0.138	&3.0(2)	&22(1) \\
\end{tabular}
\end{ruledtabular}

\caption {\label{tab2} Comparison of the calculated and observed intensities 
in NpRhGa$_{5}$ in the LT phase. For details, see the text and Table~\ref{tab1} for notation. The experimental uncertainties in parenthesis refer to the last significant digit.}
\end{table*}

Finally, we address also the question of a possible lattice distortion 
in NpRhGa$_{5}$. Onishi and Hotta~\cite{Onishi2004} examined the role of orbital degrees of freedom in determining the magnetic behavior of Np$T$Ga$_5$ compounds, on the base of a $j-j$ coupling scheme previously used to address the properties of U$T$Ga$_5$ and Pu$T$Ga$_5$ compounds~\cite{Hotta2003}. A theory based on the doublet-singlet crystal field 
model~\cite{Kiss2006} predicts a ferroquadrupolar order (FQO) in this 
compound. A homogeneous quadrupolar moment $\langle O_{xy}\rangle$ 
should be accompanied by an orthorhombic lattice distortion below $T^*$. 
The only possible allowed subgroup consistent with this FQO is $Cmmm$ 
(no. 65) corresponding to a distortion of the tetragonal $ab$-plane 
along the $[110]$ and $[1\bar{1}0]$ axes. Concomitantly, the primitive 
tetragonal lattice is converted into an orthorhombic $C$-face 
centered lattice where the orthorhombic $a$- and $b$-axes 
are the $[110]$ and $[1\bar{1}0]$ diagonals of the tetragonal unit cell, 
and the tetragonal and orthorhombic $c$-axes coincide. The 
orthorhombic distortion should split reflections of the type 
$(HKL)$ with $H, K \neq 0$. Since the magnetism of 
this material requires that the $[001]$ axis is in the scattering 
plane, because reflections have indices $(H\ K\ L{\pm}1/2)$, 
we have not performed experiments in the $(H\ K\ 0)$ plane, 
so a definitive answer as to whether this distortion appears 
at $T^*$ cannot be given.

\begin{figure}[thb]
\includegraphics*[width=0.8\columnwidth]{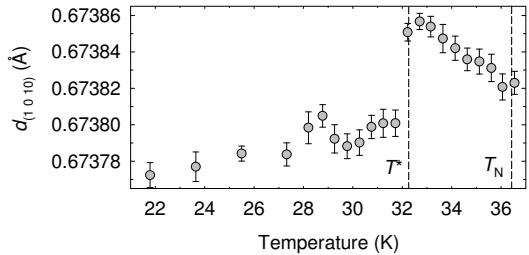}
\caption {\label{fig11} Temperature dependence of the d-spacing $d_{(1\ 0\ 10)}$
for the $\vec{Q} = (1\ 0\ 10)$ reflection in NpRhGa$_{5}$ at $B 
= 9$~T $\parallel b$-axis.
}
\end{figure}

Using a Ge(111) analyzer to improve the resolution of the spectrometer, 
we have examined reflections of the form $(H\ 0\ L)$ and 
in Fig.~\ref{fig11} we show the change of the lattice d-spacing 
as a function of temperature for the $(1\ 0\ 10)$ reflection. These 
experiments were done in the configuration with $B = 9$~T in 
an attempt to have a majority domain population, which can often 
lead to a clearer understanding of any possible lattice distortion 
associated with the magnetic ordering. The d-spacing of the studied
reflection is principally 
sensitive to the $c$-axis length.
 There is no apparent change at $\TN$ as expected for a second-order phase transition~\cite{Aoki2005a}, but a small change 
corresponding to $\Delta c/c \sim 1.3 \cdot 10^{-4}$ occurs
at $T^*$. Probably this is a magnetostrictive effect when the 
moments move away from the $[001]$ axis, but a more complete study 
with different orientation of the crystals is necessary to draw 
firm conclusions. There was no sign of any ``two-peak structure'' 
in the scans at low temperature.

\section{\label{sec:Discussion}Discussion and conclusions}

Our main objective in these studies of NpCoGa$_{5}$ and NpRhGa$_{5}$ has 
been to confirm the results of the neutron experiments on the 
same materials, and to see to what extent the RXS technique can 
bring additional information on the hybridization between the 
Np $5f$ and other electron states. From the shape of the Np 
resonances and their azimuthal dependencies there is no sign 
of any quadrupole contribution, as was found for example in NpO$_{2}$~\cite{Paixao2002}
 and UO$_{2}$~\cite{Wilkins2006}. Quadrupole ordering may, of 
course, occur at a different wave vector as compared to the magnetic 
scattering, but the shape of the resonances (Fig.~\ref{fig3}) argues that 
the moments are strictly dipole in nature. Ferroquadrupolar ordering, 
as predicted by Kiss and Kuramoto~\cite{Kiss2006} for low temperature phase of NpRhGa$_{5}$, occurs 
at the position of the charge reflections, and is thus difficult 
to observe. However, via coupling to the lattice, it should give 
rise to a lattice distortion, which we have not observed.

With respect to the resonances at the Ga $K$-edge, we have 
shown that the phenomenon of a strong resonance exists in Np 
as well as U compounds. As far as our experiments can determine, 
the effects are wholly induced by the polarization of the Np 
moments. For both the temperature dependence (Figs.~\ref{fig4} and
\ref{fig7}), 
as well as the direction, the apparent Ga dipoles (Figs.~\ref{fig6},
\ref{fig9}, and \ref{fig10}) follow that of the Np moments. 

By analyzing the polarization of the scattered photon beam and 
determining the Stokes parameters (Figs.~\ref{fig8} and \ref{fig10}), we show that 
the technique is extremely sensitive to the direction of the 
moments as well as the domain population. As shown in Fig.~\ref{fig8}, 
this information may be deduced from a single reflection even 
when a magnetic field is applied. In the case of a similar technique 
in neutron scattering ~\cite{Brown2006, Hiess2001}, the same information may be 
deduced, but the technique is restricted to zero applied field.

We have examined, briefly, the possibility of a lattice distortion 
at $T^*$ in NpRhGa$_{5}$. An effect is observed but this is probably 
a magnetostrictive change in the $c$ lattice parameter. No 
effect, at least in the $c$-axis, is observed at \TN. Our 
experience has been that when a lattice distortion occurs, at 
least \textit{two} peaks are seen with the high resolution available 
with synchrotron x-rays, and this is not the case. The FWHM of 
the lattice peak changes by only a small amount between the LT 
and HT states, suggesting strain effects rather than a lattice 
distortion.

Finally, with respect to hybridization, we have deduced the ratios 
of the dipoles at the Ga(1) site as compared to those at the 
Ga(2) site (see Fig.~\ref{fig1} for the structure). In the case of NpCoGa$_{5}$ 
the ratio of the signals Ga(1)/Ga(2) ${\sim} 7$, whereas in 
NpRhGa$_{5}$ it appears closer to about a third of this value, i.e. ${\sim} 
2.5$. Given that the nearest neighbor configuration of the Ga(1) 
has 4 equidistant Np atoms in a ferromagnetic configuration, 
compared to two Np neighbors for the Ga(2) site we would already 
expect a factor of two in the relative transferred field. In 
the case of the Co compound the distances for the Ga(1) -- Np 
and Ga(2) -- Np are 2.96 and 3.00 {\AA}, respectively. These same 
distances are equal at 2.96 {\AA} in the Rh compound. To a first 
approximation we may discount these small differences in distance 
and would expect the hybridization to be similar in the Co and 
Rh compounds, whereas it is not experimentally.

Comparisons may be made also with the hyperfine coupling constants 
as deduced in the experiments with NMR/NQR. The directions of 
the Ga dipoles can be obtained from this technique, as shown by 
Ohama \textit{et al.}\cite{Ohama2005}. For NpCoGa$_5$, 
Sakai \textit{et al.}\cite{Sakai2007} give an isotropic
hyperfine coupling constant of 46~kOe/${\mu}_\mathrm{B}$ for
Ga(1) and 36~kOe/${\mu}_\mathrm{B}$ for Ga(2). However, since 
the hybridization in the RXS case involves the polarization of 
the Ga $4p$ orbitals, which are anisotropic, it may be more 
appropriate~\cite{Kambe_priv} to compare the \textit{anisotropic} part 
of the hyperfine coupling constant, and in this case the values 
are 13~kOe/${\mu}_\mathrm{B}$ for Ga(1) and 4~kOe/${\mu}_\mathrm{B}$ for Ga(2).
This ratio is in rough accord 
with our measurements, since the RXS result should be divided 
by two to normalize for the number of Np nearest neighbors. No 
NMR/NQR experiments have yet been reported on NpRhGa$_{5}$. So far, 
no calculations have been performed of the possible signals at 
the Ga $K$-edges in these 115 materials, but there seems sufficient 
information at this stage, together with the previous work on 
U 115 compounds~\cite{Kuzushita2006}, to encourage such theoretical 
efforts. Both the shapes of the resonances, which are different 
between NpCoGa$_{5}$ and NpRhGa$_{5}$ (see discussion connected to Fig.~\ref{fig3}),
 and the differing ratios of the Ga(1)/Ga(2) dipole strengths 
suggest that the transition-metal $d$ states play a significant 
role in the physics of these compounds and do not simply act 
as ``spacers''.

\begin{acknowledgments}
We thank the safety personnel at the ESRF for their help in running 
these experiments, C. Detlefs for discussions and critical reading of the manuscript and S. Kambe and R. Walstedt for correspondence 
concerning the NMR/NQR experimental results. B.D. and S.B.W. 
thank the European Commission for support in the frame of the
``Training and Mobility of Researchers'' 
program. The high purity neptunium metal required for the fabrication of these compounds was made available through a loan agreement between Lawrence Livermore National Laboratory and ITU, in the frame of a collaboration involving LLNL, Los Alamos National Laboratory, and the U.S. Department of Energy.
F.H. would like to thank for the financial support by the Grant-in-Aid for Scientific Research from the Japanese Ministry of Education, Culture, Sports, Science and Technology (No. 18740219) and by the REIMEI Research Resources of Japan Atomic Energy Agency.

\end{acknowledgments}

\bibliography{d:/blanka/diff_papers/ref_database}

\end{document}